# Silicon Micro-Disk Resonator Crossbar Array for High-Speed and High-Density Photonic Convolution Processing

Long Huang[1] and Jianping Yao[1*]

[1]*Microwave Photonics Research Laboratory,*
*School of Electrical Engineering and Computer Science, University of Ottawa,*
*Ottawa, Ontario K1N 6N5, Canada*
**Corresponding author: jpyao@uottawa.ca*

**Abstract:** Advanced artificial intelligence (AI) algorithms, particularly those based on artificial neural networks, have garnered significant attention for their potential applications in areas such as image recognition and natural language processing. Notably, neural networks make heavy use of matrix-vector multiplication (MVM) operations, causing substantial computing burden on existing electronic computing systems. Optical computing has attracted considerable attention that can perform optical-domain MVM at an ultra-high speed. In this paper, we introduce a novel silicon photonic micro-disk resonator (MDR) crossbar signal processor designed to support matrix-vector multiplication (MVM) with both high processing speed and enhanced computational density. The key innovation of the proposed MDR crossbar processor is the placement of two MDRs at each crosspoint, enabling simultaneous routing and weighting functions. This design effectively doubles the computational density, improving overall performance. We fabricate a silicon photonic MDR crossbar processor, which is employed to perform convolutional tasks in a convolutional neural network (CNN). The experimental results demonstrate that the photonic processor achieves a classification accuracy of 96% on the MNIST dataset. Additionally, it is capable of scaling to a computational speed of up to 160 tera-operations per second (TOPS) and a computational density as high as 25.6 TOPS/mm². Our approach holds significant promise for enabling highly efficient, scalable on-chip optical computing, with broad potential applications in AI and beyond.

Artificial neural networks (ANNs) have demonstrated superior performance in a variety of artificial intelligence (AI) tasks, such as image classification [1] and language translation [2]. However, ANNs require computationally expensive matrix-vector multiplication (MVM) operations, which becomes challenging for traditional von Neumann computing schemes owing to speed and energy inefficiency. To address this issue, substantial efforts have been devoted to the hardware design of neuromorphic computing frameworks aimed at enhancing the computational speed of neural networks. These efforts include advancements in both electronic architectures [3-5] and optical architectures [6, 7]. Among extensive neuromorphic computing systems, optical matrix-vector multiplication (MVM) processors distinguish themselves by high bandwidth, low latency, and low energy consumption [8]. Beyond conventional image classification tasks, an optical MVM processor is better suited for integration with optical systems, such as communication links [9-11]. Because of these advantages, a few optical MVM processing approaches have been proposed. For example, diffractive neural networks have been proposed in which MVM operations are performed based on layers of diffractive elements [12-16]. However, diffractive neural networks rely on bulky optical diffraction elements, making the system large and costly. Photonic integrated chips have been also used to perform MVM processing. Based on matrix singular value decomposition (SVD) and unitary matrix parametrization, Shen et al. [17] designed and fabricated a fully optical neural network based on Mach-Zehnder interferometer (MZI) arrays. Since then, MZI array has become a widely adopted architecture for optical MVM processing [18-21]. Aside from the MZI array architecture, other architectures for the MVM operations have been proposed. One such architecture, the crossbar, was initially introduced in the electronic computing field [22], and later adapted for the optical domain. In [23], an optical crossbar array combined with phase change material (PCM) was proposed. In this structure, the matrix weights are controlled by tuning the PCM modulators. However, it is different from a regular phase modulator such as a $LiNbO_3$ phase modulator which is electronically controlled, a PCM modulator is adjusted optically, which makes the system complicated. In [24], an optical crossbar array combined with micro-ring resonators was proposed. In this structure, the weights are adjusted by thermally tuning the micro-resonators. Compared to a MZI array, a micro-resonator array has a more compact size. For example, the length of a MZI in [17] is 100 μm while the radius of a micro-resonator is only 10 μm [24]. For all the optical crossbar arrays reported [22-24], at one cross only one weight is implemented, the computational density is low.

In this paper, we propose a novel optical crossbar architecture based on micro-disk resonators (MDRs) to improve computational speed and computational density. The key innovation of the proposed MDR crossbar array processor is the placement of two MDRs at each cross, allowing them to simultaneously perform routing and weighting functions. This dual functionality effectively doubles the computational density. Moreover, compared to micro-ring resonators, MDRs offer lower loss, making them a more efficient choice for optical processing [25]. We fabricate a silicon photonic MDR crossbar array and experimentally demonstrate its capability in an image classification task. Specifically, we perform classification on 50 images from the MNIST dataset. The experimental results show that the photonic processor achieves a classification accuracy of 96%, with a computational speed being able to scalable up to 160 tera-operations per second (TOPS) and a computational density reaching as high as 25.6 TOPS/mm². The proposed architecture demonstrates strong potential for the on-chip realization of matrix-vector multiplication (MVM) with enhanced computational density, higher processing speed, and lower power consumption, paving the way for the next generation of AI platforms.

# Results

### The Signal Processor Architecture

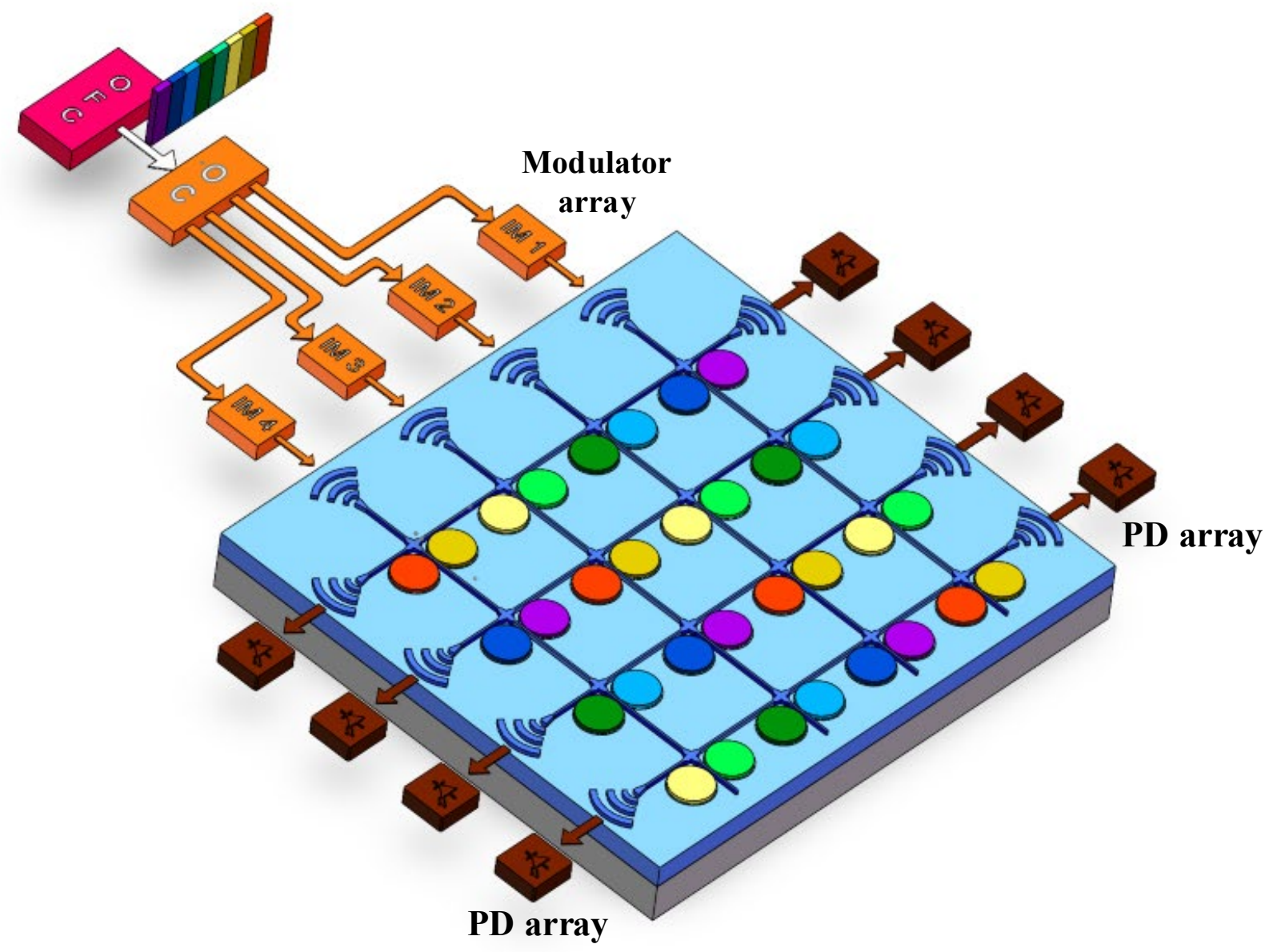

Fig. 1. Schematic diagram of the proposed MDR crossbar signal processor for MVM processing. OFC: optical frequency comb source, OC: optical coupler, IM: intensity modulator.

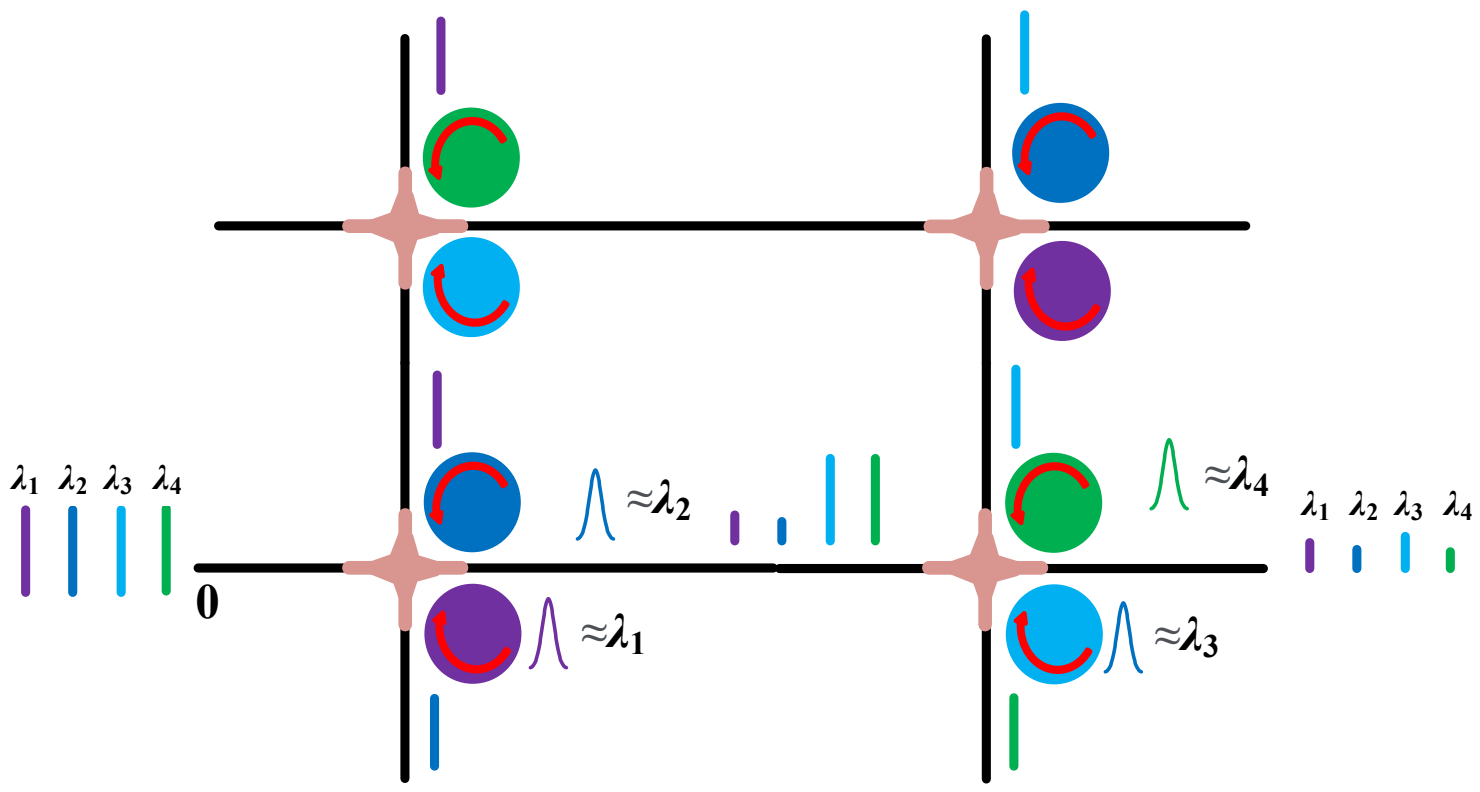

Fig. 2. The routing and weighting function of the MDRs.

Fig. 1 shows the schematic diagram of the proposed MDR array processor implemented on a silicon photonic platform. As can be seen the processor has a two-dimensional 4×4 crossbar structure consisting of 16 crossbar with two MDRs at each cross of a crossbar. An optical frequency comb (OFC) generated by a comb source is sent to an optical coupler (OC) to split the comb light into 4 channels, which are sent to four intensity modulators (IMs) driven by an input vector signal. Then, the optical signals after the IMs are sent to the MDR crossbar array. Two thermally tunable MDRs are placed at a cross of a crossbar, as shown in Fig. 1. The MDRs are used for implementing routing and weighting. The principle is schematically shown in Fig. 2. Suppose four wavelengths, denoted by $\lambda_1$, $\lambda_2$, $\lambda_3$, $\lambda_4$, are sent to port 0, the two MDRs placed at the lower left cross are intentionally employed to perform routing and weighting for $\lambda_1$ and $\lambda_2$, so the resonance wavelengths of the two MDRs are adjusted to be approximately at $\lambda_1$ and $\lambda_2$. As a result, $\lambda_1$ is routed to the upward vertical direction while $\lambda_2$ is routed to the downward vertical direction. By adjusting the difference between the resonance wavelengths of the MDRs and the OFC wavelengths, the weighting to the two wavelengths can be achieved. Then, the residual wavelengths at $\lambda_1$ and $\lambda_2$, along with the intact wavelengths at $\lambda_3$ and $\lambda_4$, will proceed to the lower right cross. At the right cross, the two MDRs at the crossbar are assigned to route and weight the intact wavelengths at $\lambda_3$ and $\lambda_4$, and the routing and weighting process being the same as the previous cross. On the other hand, the wavelength at $\lambda_1$ that is routed upwards is directed to the upper left cross. If the resonance wavelengths of the two MDRs at the left upper cross is not tuned to be near $\lambda_1$, the upper cross is transparent for $\lambda_1$, and $\lambda_1$ can go through the upper cross without experiencing a loss. For $\lambda_2$ routed to the downward vertical direction, the same principle applies.

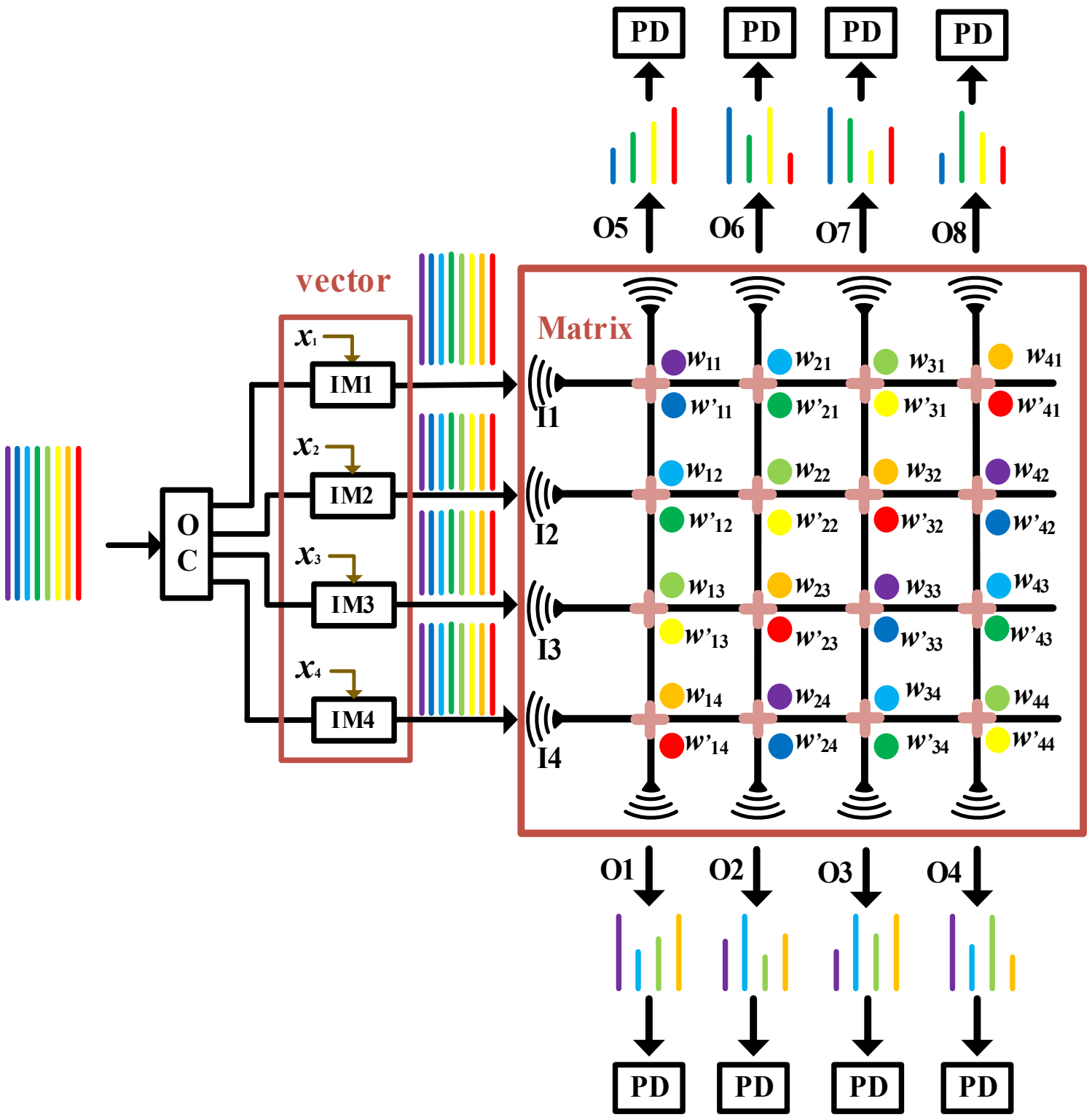

Fig. 3. Wavelengths and weights arrangement of the crossbar array processor.

One arrangement of the wavelengths and weights is given in Fig. 3. At each output port denoted by O1 to O8, the routed and weighted wavelengths are combined. The optical signal at each output port is sent to a photodetector (PD) for optical-to-electrical conversion and summing operation simultaneously. As shown in Fig. 3, the input signals are denoted by $x_1$ to $x_4$, and the output signals are denoted by $y_1$ to $y_4$ and $y'_1$ to $y'_4$. The two MVM operations achieved by the proposed MDR crossbar array processor are mathematically given by

$$\begin{bmatrix} y_1 \\ y_2 \\ y_3 \\ y_4 \end{bmatrix} = \begin{bmatrix} w_{11} & w_{12} & w_{13} & w_{14} \\ w_{21} & w_{22} & w_{23} & w_{24} \\ w_{31} & w_{32} & w_{33} & w_{34} \\ w_{41} & w_{42} & w_{43} & w_{44} \end{bmatrix} \begin{bmatrix} x_1 \\ x_2 \\ x_3 \\ x_4 \end{bmatrix} \tag{1a}$$

$$\begin{bmatrix} y'_1 \\ y'_2 \\ y'_3 \\ y'_4 \end{bmatrix} = \begin{bmatrix} w'_{11} & w'_{12} & w'_{13} & w'_{14} \\ w'_{21} & w'_{22} & w'_{23} & w'_{24} \\ w'_{31} & w'_{32} & w'_{33} & w'_{34} \\ w'_{41} & w'_{42} & w'_{43} & w'_{44} \end{bmatrix} \begin{bmatrix} x_1 \\ x_2 \\ x_3 \\ x_4 \end{bmatrix} \tag{1b}$$

As can been, compared to the previous electronic crossbar or the optical crossbar [22-24] which can only achieve one weight at each cross, the proposed MDR crossbar can achieve two weights at each cross, leading to doubling of the computational density.

## The chip

The silicon photonic MDR crossbar has a two-dimensional crossbar structure with four input and eight output ports. The photo of the fabricated MDR crossbar is shown in Fig. 4. The cell of the crossbar has a square shape having a length of 75 μm, and a low-loss waveguide cross based on 1×1 multimode interferometer (MMI) is employed at the waveguide intersection to enable low-crosstalk horizontal and vertical optical transmission. To reduce the sidewall roughness and to increase the light confinement capacity and optical coupling between the bus waveguide and the disk, each MDR is designed such that an additional slab waveguide is used to wrap the disk and the bus waveguide. The parameters of the MDRs in the cell are as follows: the disk radius is 6.4 μm and the disk height is 220 nm. The height of the bus waveguides is also selected to be 220 nm and the height of the additional slab waveguide is selected to be 90 nm. Note that the widths of the slab waveguide wrapping the disk and the bus waveguide are controlled to be 200 nm. To enable effective coupling, the two slab waveguides in the coupling region are designed such that they fully overlap. The bus waveguide has a wider width of 500 nm, to ensure an effective excitation of the first-order whispering gallery-mode (WGM). To enable thermal tuning, each MDR has a high-resistivity metallic micro-heater as a phase shifter, placed on top of the disk. The illustration of one cross is shown in Fig. 5(a), and the cross-sectional view of the planar structure of two MDRs along the dash line presented in Fig. 5(a) is shown in Fig. 5(b). The MMI-based cross consists of two orthogonal intersecting MMI waveguides with each having a width of 1.44 μm and a length of 6 μm, to enable low-loss and low-crosstalk transmission.

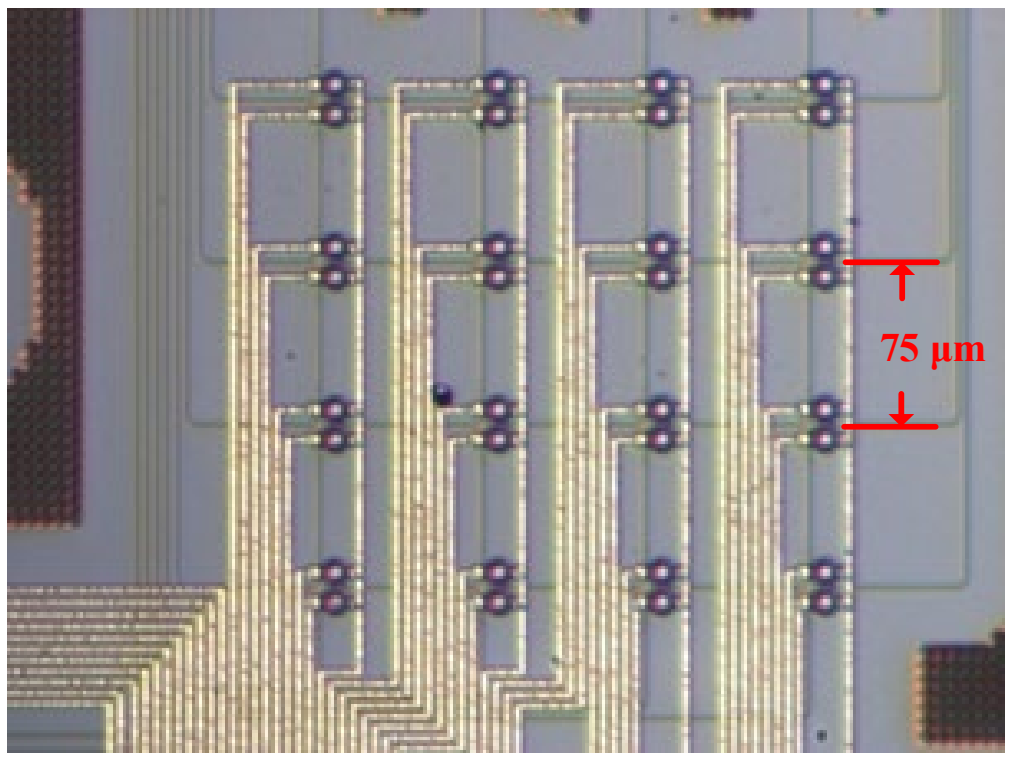

Fig. 4. Photo of the fabricated MDR crossbar.

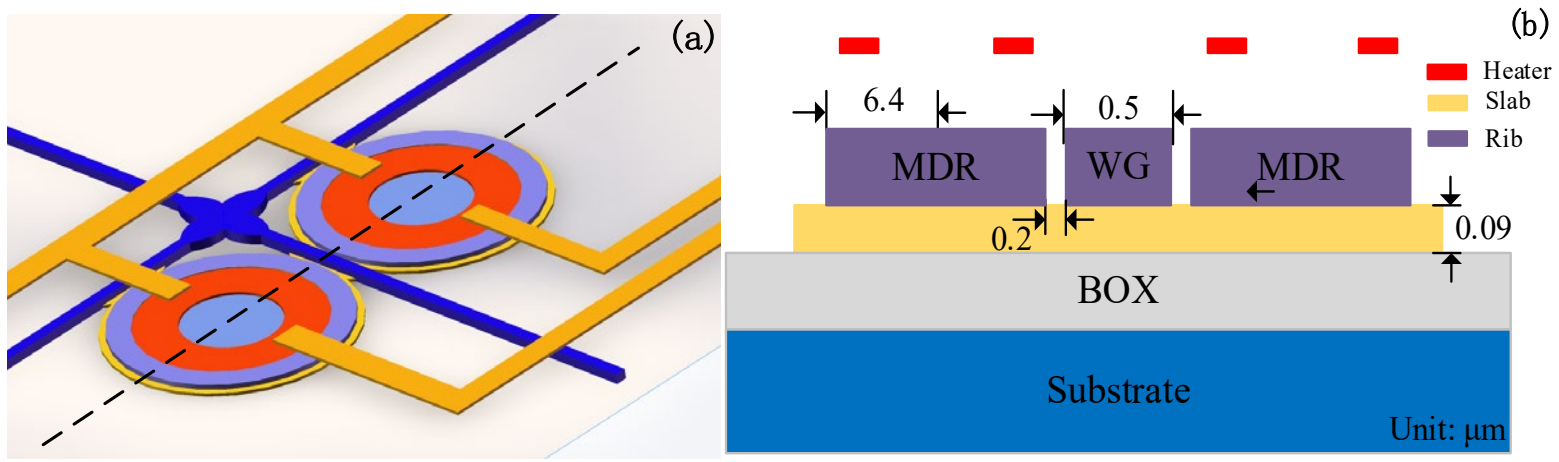

Fig. 5. (a) Illustration of two MDRs at one cross. (b) Cross-sectional view of the planar structure of the two MDRs. MDR: micro-disk resonator, WG: waveguide, BOX: buried oxide.

## Experiment

Based on the fabricated MDR crossbar array processor, we experimentally demonstrate the optical convolution operation using the processor with 2×2 kernels. The experiment setup is shown in Fig. 6. We use the 8 MDRs in the box, as shown in Fig. 6, for a proof-of-concept experiment.

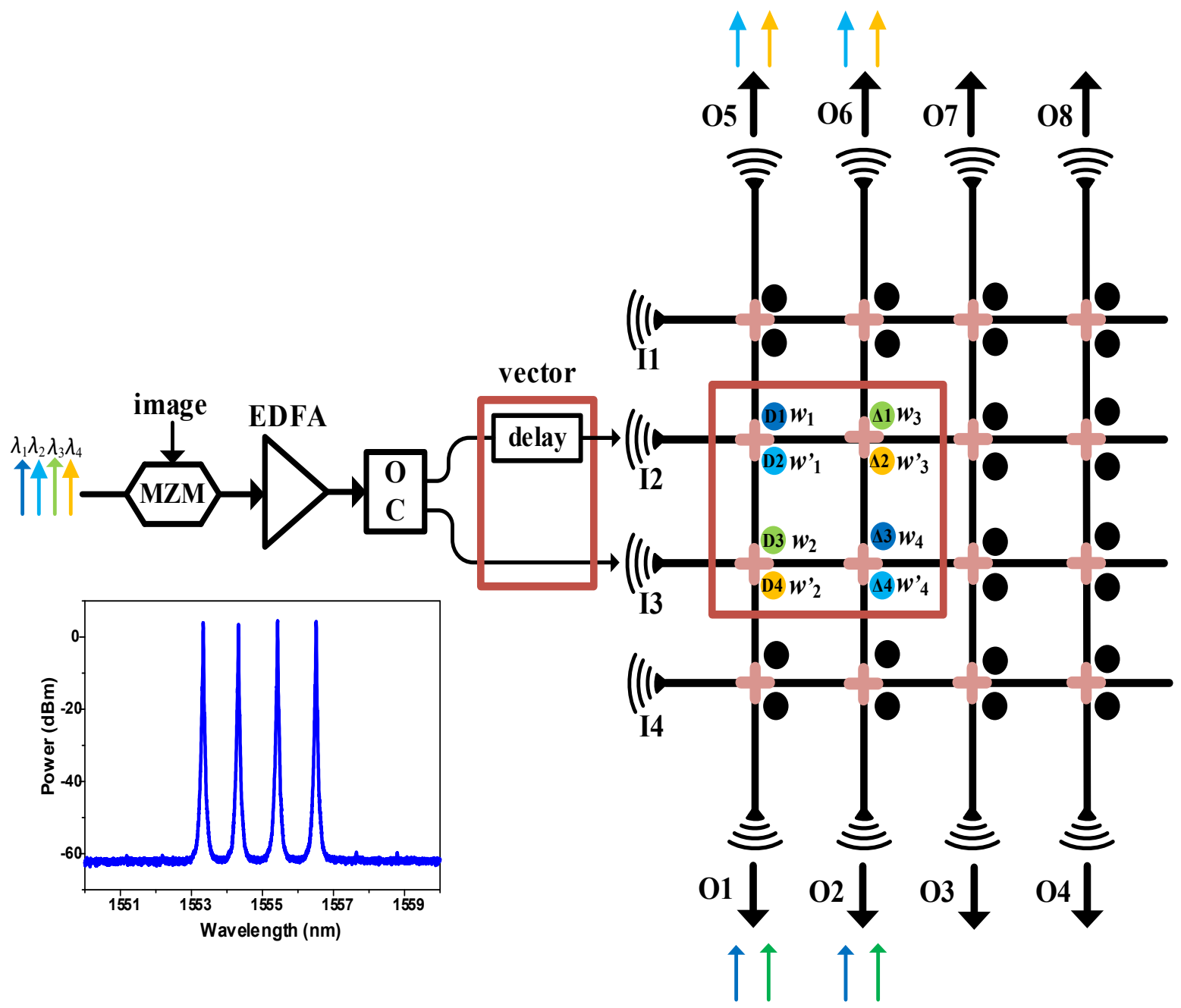

Fig. 6. Experimental setup of the convolution with 2×2 kernels. Inset: the optical spectrum of the light from the OFC.

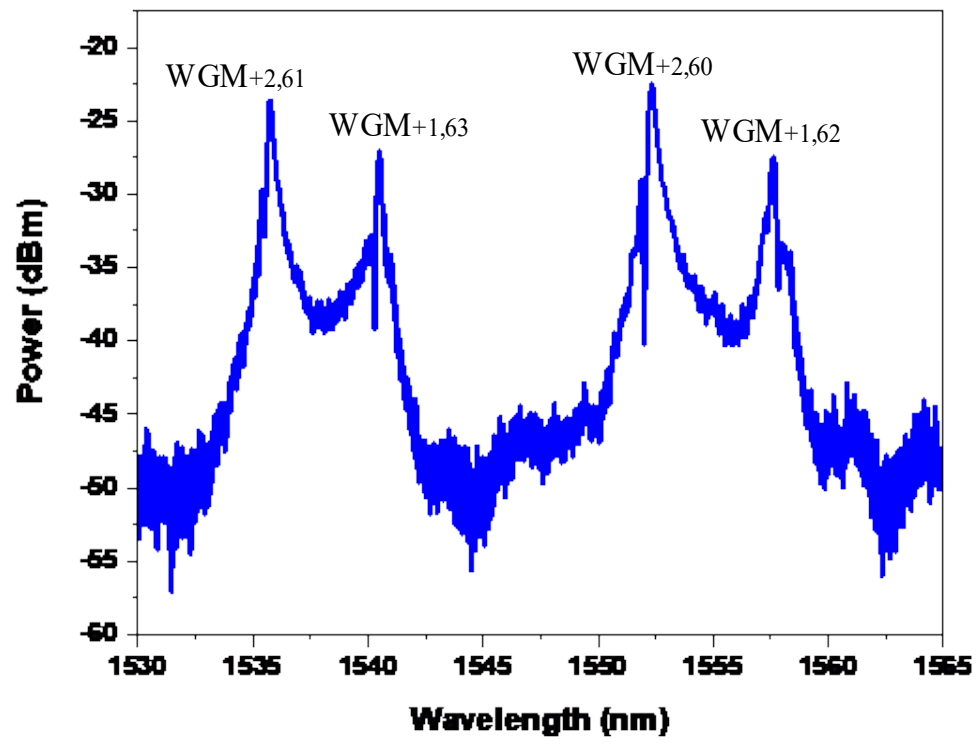

Fig. 7. Measured transmission spectrum from O1 to O5 without applying voltages to the MDR.

We measure the reflection spectrum of one MDR by connecting the ports of an optical vector analyzer (OVA) to O5 and I1, and the measured spectrum is shown in Fig. 7. The peaks in the spectrum represents WGM$p$,$q$ resonance in which $p$ and $q$ are the radial and azimuthal harmonic numbers. The periodicity of the peaks tells that first-order and second-order WGMs are effectively excited in the disks. The first-order WGM is measured to have a free spectral range (FSR) of 16.6 nm, and the second-order WGM has an FSR of 17.1 nm.

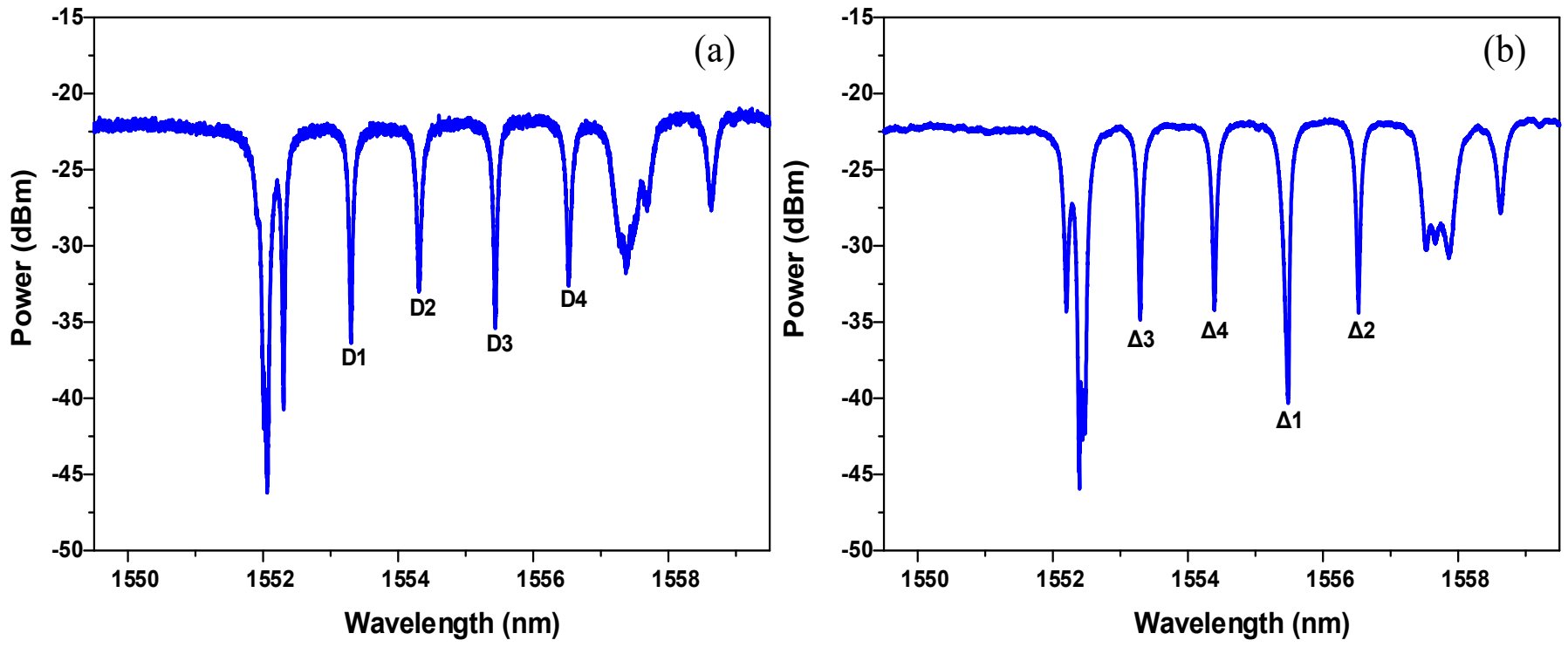

Fig. 8. Measured transmission spectrum from (a) O1 to O5, and from (b) O2 to O6.

Then we apply a voltage to tune the resonance wavelength of each MDR. The utilized MDRs are labeled by D1 to D4 and Δ1 to Δ4 as shown in Fig. 6. The resonance wavelength of D1 is tuned to be $\lambda_1$; that of D2 is tuned to be $\lambda_2$; that of D3 is tuned to be $\lambda_3$; that of D4 is tuned to be $\lambda_4$. On the other hand, the resonance wavelength of Δ1 is tuned to be $\lambda_3$; that of Δ2 is tuned to be $\lambda_4$; that of Δ3 is tuned to be $\lambda_1$; that of Δ4 is tuned to be $\lambda_2$. We also keep the voltages to the other MDRs to be zero. Then, the transmission spectrum from O1 to O5 is measured and shown in Fig. 8(a) while the transmission spectrum from O2 to O6 is measured and shown in Fig. 8(b). The notches and their corresponding MDRs are labeled in Fig. 8.

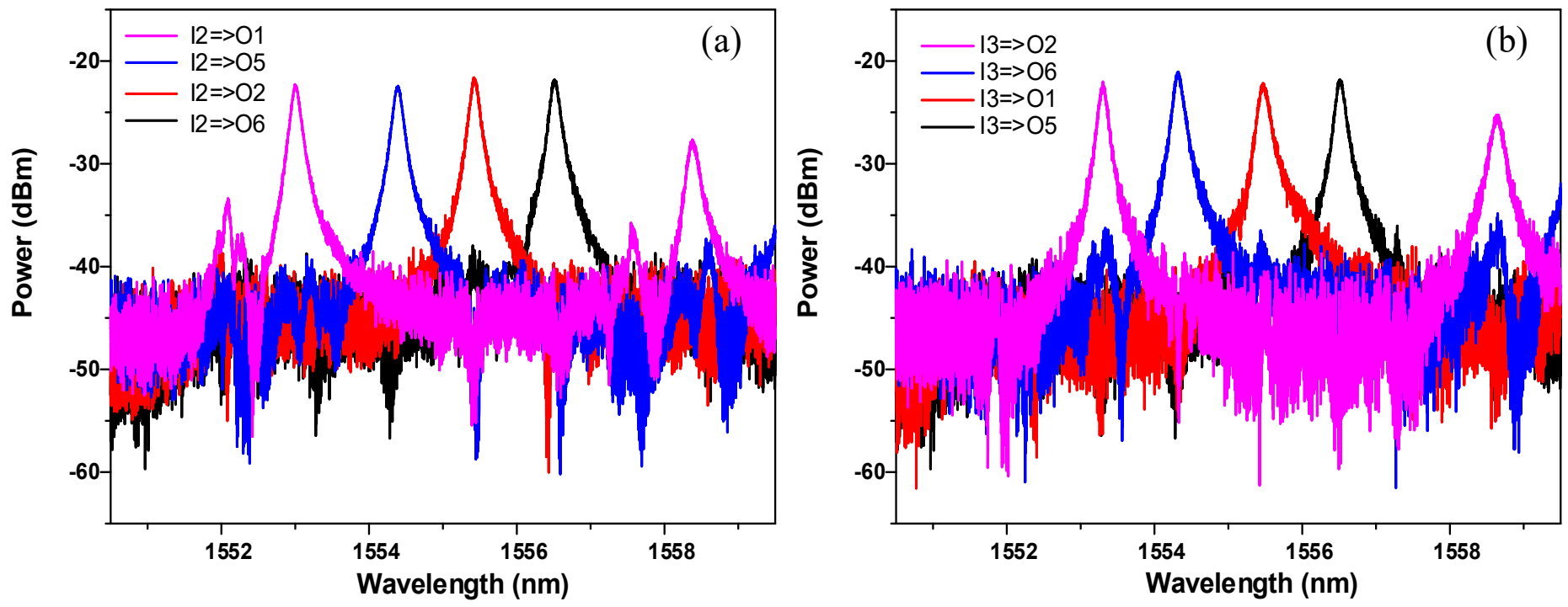

Fig. 9. Measured transmission spectrum. (a) From I2 to O1, O5, O2, O6, (b) from I3 to O2, O6, O1, O5.

Then, we measure the transmission spectrum from I2 to O1, O2, O5 and O6, and the spectrum is shown in Fig. 9(a). We measure the transmission spectrum from I3 to O1, O2, O5 and O6, and the spectrum is shown in Fig. 9(b). As can be seen from Fig. 9(a), if the OFC is sent to I2 port, $\lambda_1$ is routed to O1, $\lambda_2$ is routed to O5, $\lambda_3$ is routed to O2, and $\lambda_4$ is routed to O6, respectively. We can also see from Fig. 9(b), if the OFC is sent to I3 port, $\lambda_1$ is routed to O2, $\lambda_2$ is routed to O6, $\lambda_3$ is routed to O1, and $\lambda_4$ is routed to O5, respectively.

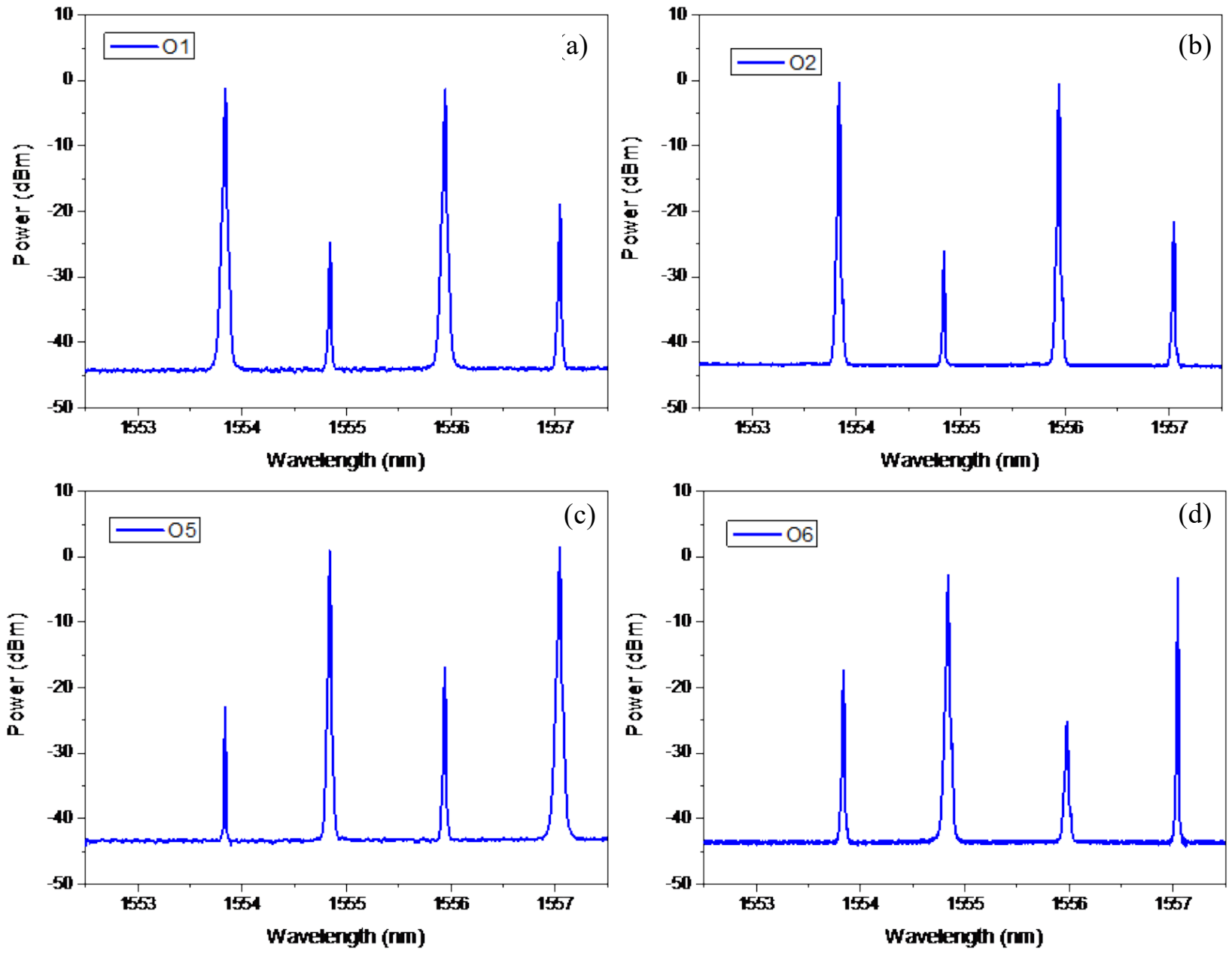

Fig. 10. Measured spectrum when the OFC is sent to both I2 and I3 ports. (a) The output of O1 port, (b) the output of O2 port, (c) the output of O5 port, and (d) the output of O6 port.

Then, the OFC is sent to both I2 and I3 ports. The optical spectra at O1, O2, O5 and O6 ports are shown in Fig. 10. As can be seen, $\lambda_1$ and $\lambda_3$ are routed to the output ports O1 and O2 while $\lambda_2$ and $\lambda_4$ are routed to the output ports O5 and O6.

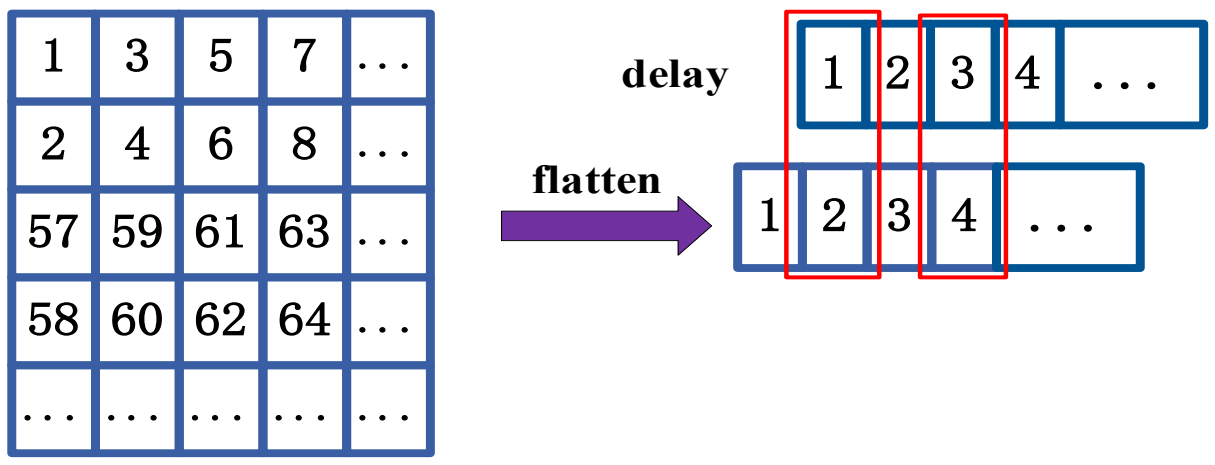

Fig. 11. A 2D image is flattened to a 1D signal and the generation of two inputs by adding a delay.

Finally, the OFC is sent to a Mach-Zehnder modulator (MZM) that is biased at the quadrature point. The 2D image is preprocessed to act as the input vector. The main part of the preprocessing is to flatten the 2D image into a 1D signal. The schematic diagram of the flattening operation is shown in Fig. 11, and the detailed preprocessing operation is given in the Supplementary Note 1. The generated 1D signal is used to drive the MZM for electrical-to-optical conversion. The clock frequency of the 1D signal in the experiment is 8 GHz. Then, the optical signal at the output of the MZM is split into two branches by an optical coupler (OC). One of the two branches is delayed by one symbol duration to generate $2\times1$ vectors as also shown in Fig. 11. Since we use one intensity modulator to simulate the generation of $2\times1$ vectors, every other symbol is effective as framed by the red box shown in Fig. 11. Two convolution kernels are loaded to the MDR crossbars as shown Fig. 6. Suppose the two kernels are $[w_1, w_2; w_3, w_4]$ and $[w_1', w_2'; w_3', w_4']$. At each output port, we can achieve $2\times1$ dot product. Therefore, by using two output ports, we can achieve the $2\times1$ vector and $2\times2$ matrix multiplication. To achieve both positive and negative weights, a calibration process should be implemented, and the detail of the calibration process is given in Supplementary Note 1.

Firstly, two convolution kernels [1,1;1,1] and [1,1;-1,-1] are used to compute the convolution by the MDR crossbar array processor. Specifically, $w_1$ = 1, $w_2$ = 1, $w_3$ = 1, $w_4$ = 1, $w_1'$ = 1, $w_2'$ = 0, $w_3'$ = 1, and $w_4'$ = 0 are loaded to the MDR crossbar. Four images are randomly selected from the MNIST dataset as the input vector. The selected MNIST images are shown in Fig. 12(a). The selected MNIST images are flattened to be 1-D signals which are then normalized with zero mean, the preprocessed waveforms are generated by an arbitrary waveform generator (AWG), as shown in Fig. 12(b). The convolution results of the MDR crossbar and a digital computer for the kernel [1,1;1,1] are shown in Fig. 13(a). The normalized mean square errors (NMSEs) of the four images between the photonic processor and the digital computer are 0.0250, 0.0228, 0.0237 and 0.0228. Then, the flatten convolution results are then reconverted to 2D images, as shown in Fig. 13(b). Furthermore, the convolution results of the MDR crossbar processor and the digital computer for the kernel [1,1;-1,-1] are shown in Fig. 14(a). The NMSEs of the four images between the optical processor and the digital computer are 0.1316, 0.1723, 0.1268 and 0.1639. The flatten convolution results are reconverted to 2D images, as shown in Fig. 14(b).

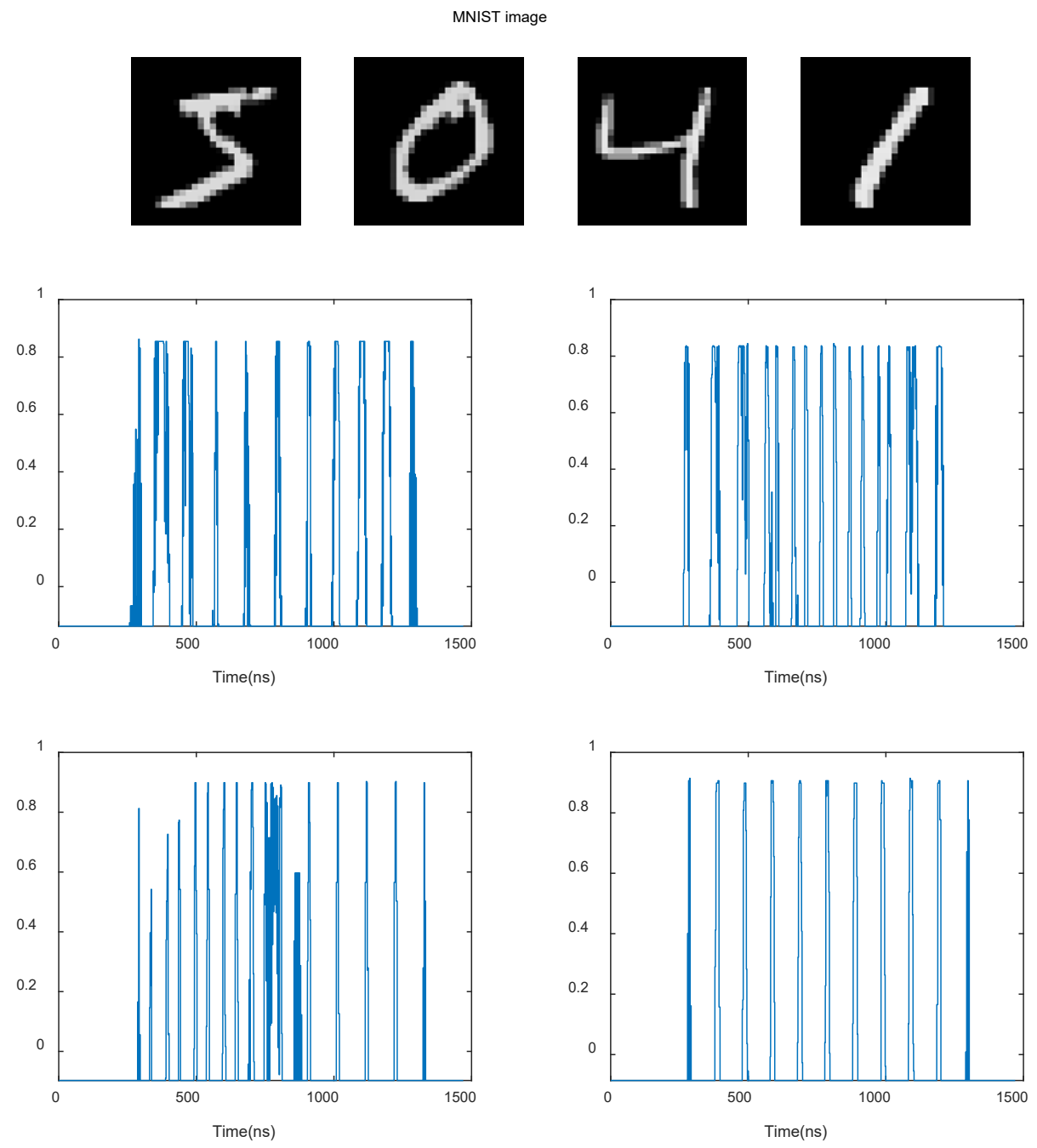

Fig. 12. (a) Four selected MNIST images, (b) 1D waveforms generated from the four images.

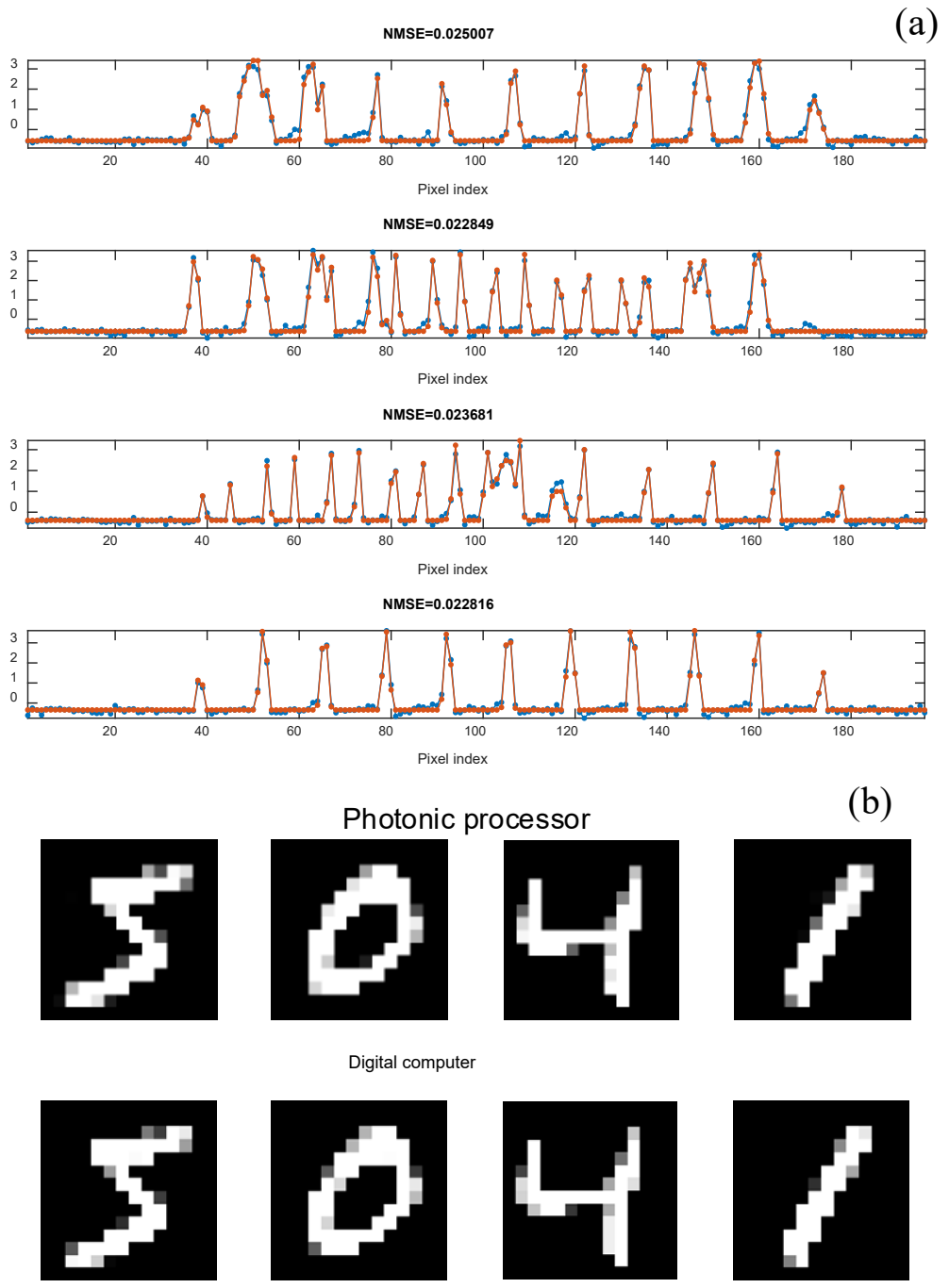

Fig. 13. The convolution with the kernel [1,1;1,1] computed by the optical MDR crossbar processor and a digital computer. (a) The 1D result, and (b) the 2D result.

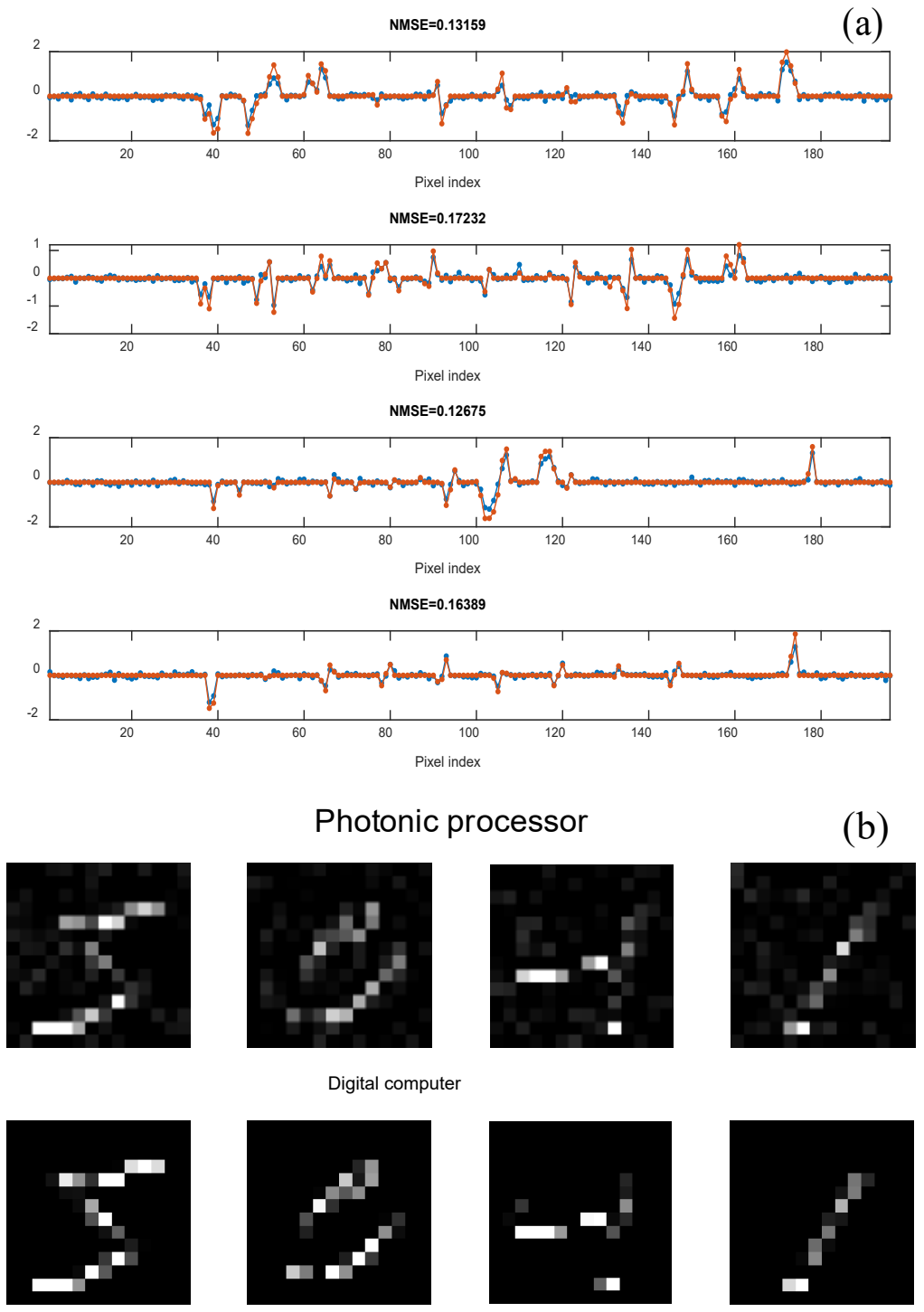

Fig. 14. The convolution with the kernel [1,1; -1,-1] computed by the optical MDR crossbar processor and a digital computer. (a) The 1D result, and (b) the 2D result.

Then, the convolution kernels [1,0;0,1] and [-1,-1;1,1] are used to calculate the convolution using the MDR crossbar array processor. Specifically, $w_1 = 1$, $w_2 = 0.5$, $w_3 = 0.5$, $w_4 = 1$, $w'_1$=0, $w'_2$=0, $w'_3$=1, and $w'_4$=1 are loaded to the MDR crossbar. The convolution results of the MDR crossbar and the digital computer for the kernel [1,0; 0,1] shown in Fig. 15(a). The NMSEs of the four images between the optical processor and the digital computer are 0.0714, 0.0432, 0.0757 and 0.0640. Then the flatten convolution results are reconverted to 2D images, as shown in Fig. 15(b). On the other hand, The convolution results of the MDR crossbar and the digital computer for the kernel [-1,-1;1,1] are shown in Fig. 16(a). The NMSEs of the four images between the optical processor and the digital processor are 0.0918, 0.0830, 0.0862 and 0.1089. Then the flatten convolution results are reconverted to 2D images, as shown in Fig. 16(b).

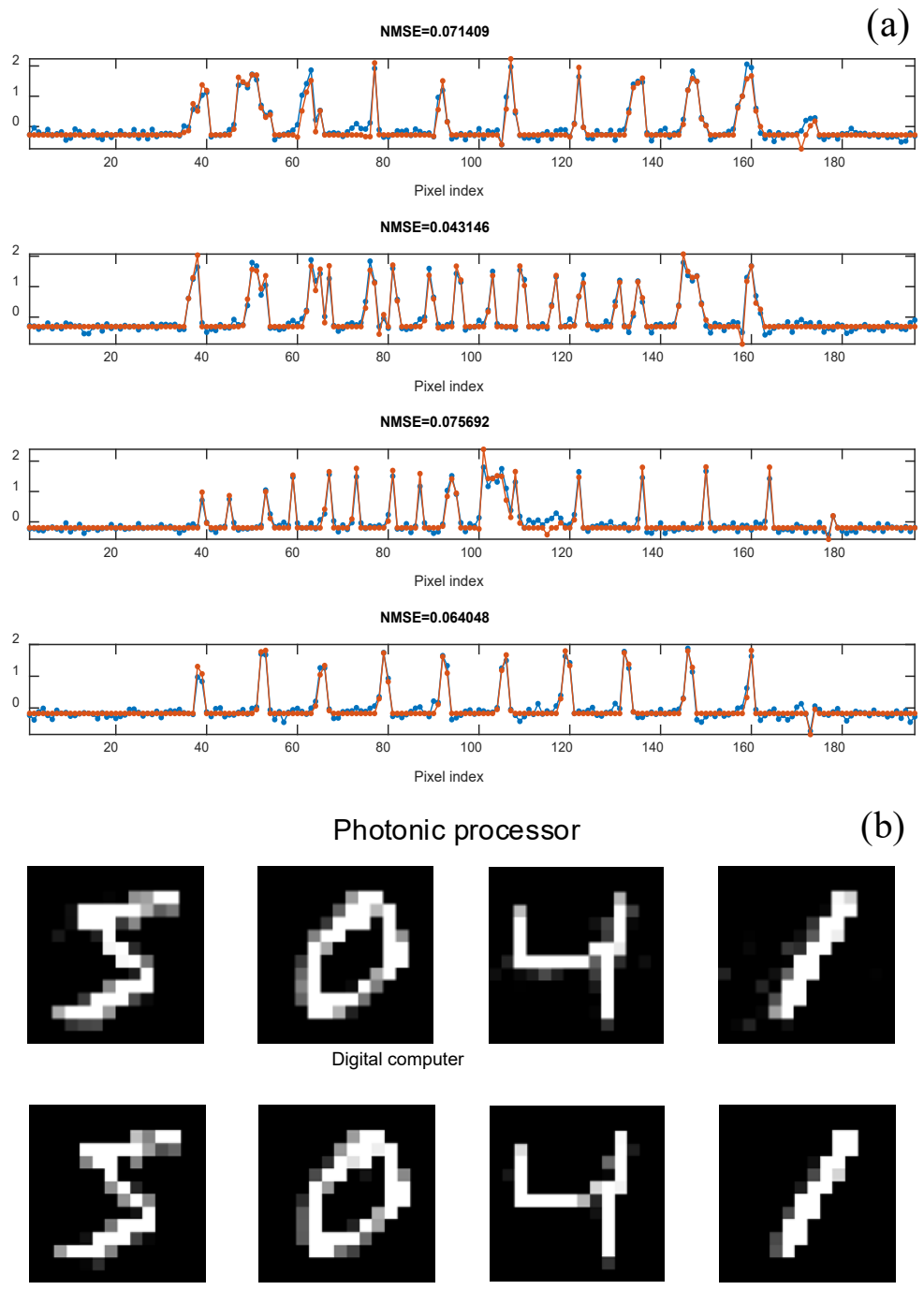

Fig. 15. The convolution with the kernel [1,0;0,1] computed by the optical MDR crossbar processor and a digital computer. (a) The 1D result, and (b) the 2D result.

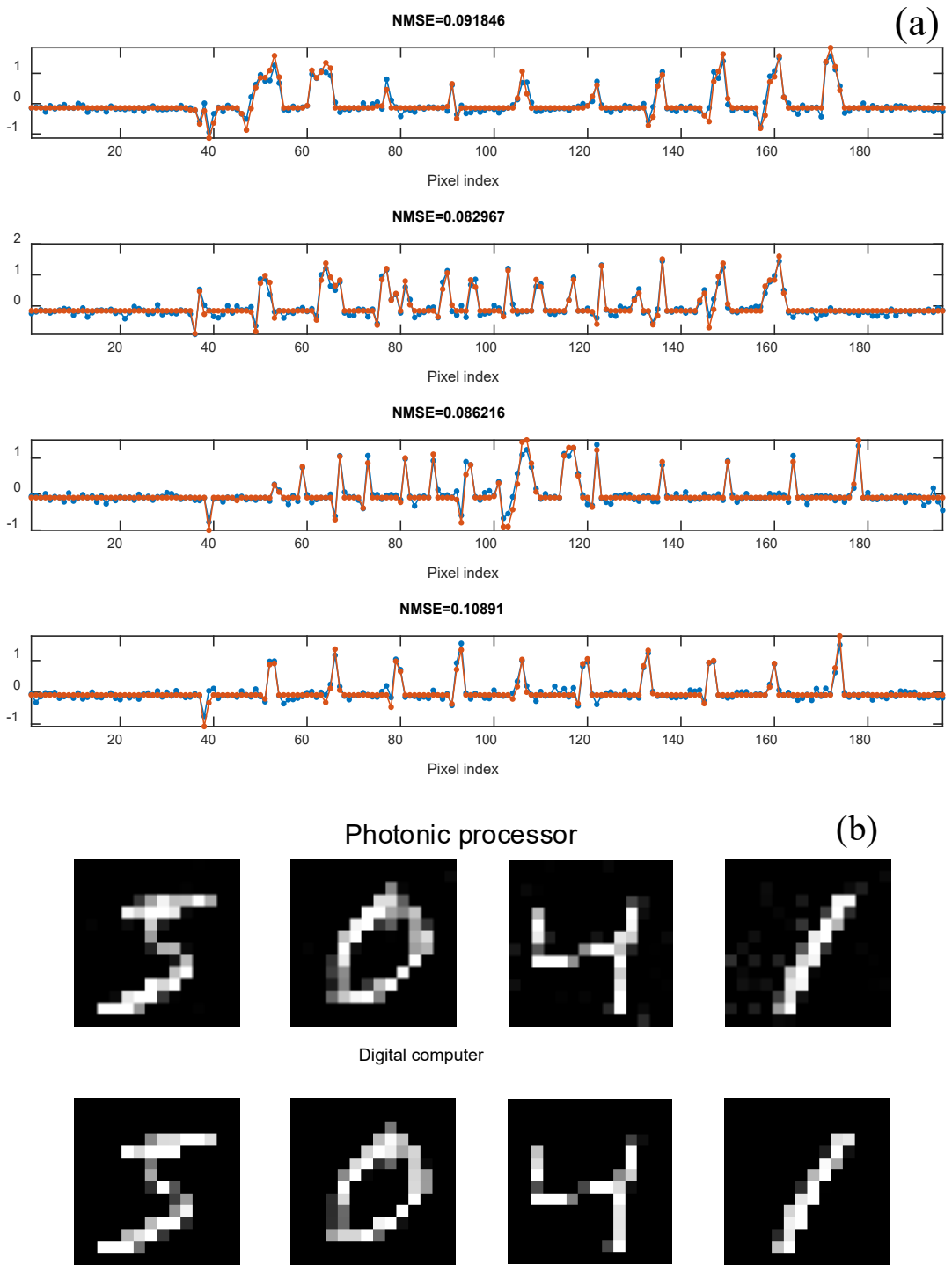

Fig. 16. The convolution with the kernel [-1,-1; 1,1] computed by the optical MDR crossbar processor and a digital computer. (a) The 1D result, and (b) the 2D result.

The bit precision of MVM operations with the photonic processor is also calculated. The standard deviation of the MVM results calculated by the photonic processor is 0.0248, resulting in a bit precision of 5.3-bit (more details about bit precision can be seen in Supplementary Note 3).

The proposed photonic processor is employed to implement a convolutional neural network (CNN) for the classification of images from the MNIST dataset. The structure of the CNN model is shown Fig. 17. The convolutional layer in the CNN model has two channels and the kernel size is 2×2. The two kernels are achieved in the photonic processor, generating two 1×196 feature maps. After being activated using the ReLU nonlinear function, two 1×196 feature maps are flattened into a 1×392 vector and then fed to a fully connected (FC) layer with the output dimension being 100. The output of the FC layer is activated by the ReLU nonlinear function, and sent to another FC layer with the output dimension being 10. The output of the second FC layer is sent to a SoftMax layer to generate the output of the CNN model. The weights of the CNN model are trained offline using the back propagation algorithm. A total of 50000 images from the MNIST are used to train the CNN model, and the trained accuracy is 96.61% after 10 epochs. The training accuracy and the cross-entropy loss during the 10 epochs are shown in Fig. 18(a). The offline trained kernels are [-0.3576, -0.5353; 0.7799, -0.5282] and [-0.0244, -0.4925; -0.8154, -0.0554]. In the inference stage, the kernels are loaded to the MDR crossbar array to classify the MNIST dataset. The kernels are also used by the digital computer as reference. A total of 50 images from the MINIST dataset are experimentally tested, and the confusion matrix of the digital computer and the photonic processor is shown in Fig. 18(b). As illustrated by the experimental results, the photonic processor and the digital computer have the same classification results, which verifies the effectiveness of the proposed photonic processor. In the experiment, both the photonic processor and the digital computer have 2 mis-classified images leading to an accuracy of 96%. The results of the classification of the 50 images from the MNIST dataset by the digital computer and the photonic processor are given in Supplementary Note 2. The discrepancy between results from the digital computer and the photonic processor is mainly caused by the limited bit precision which is caused by a few factors, including the electrical and optical noise and instability of some optical devices (polarization state jitter, temperature drift).

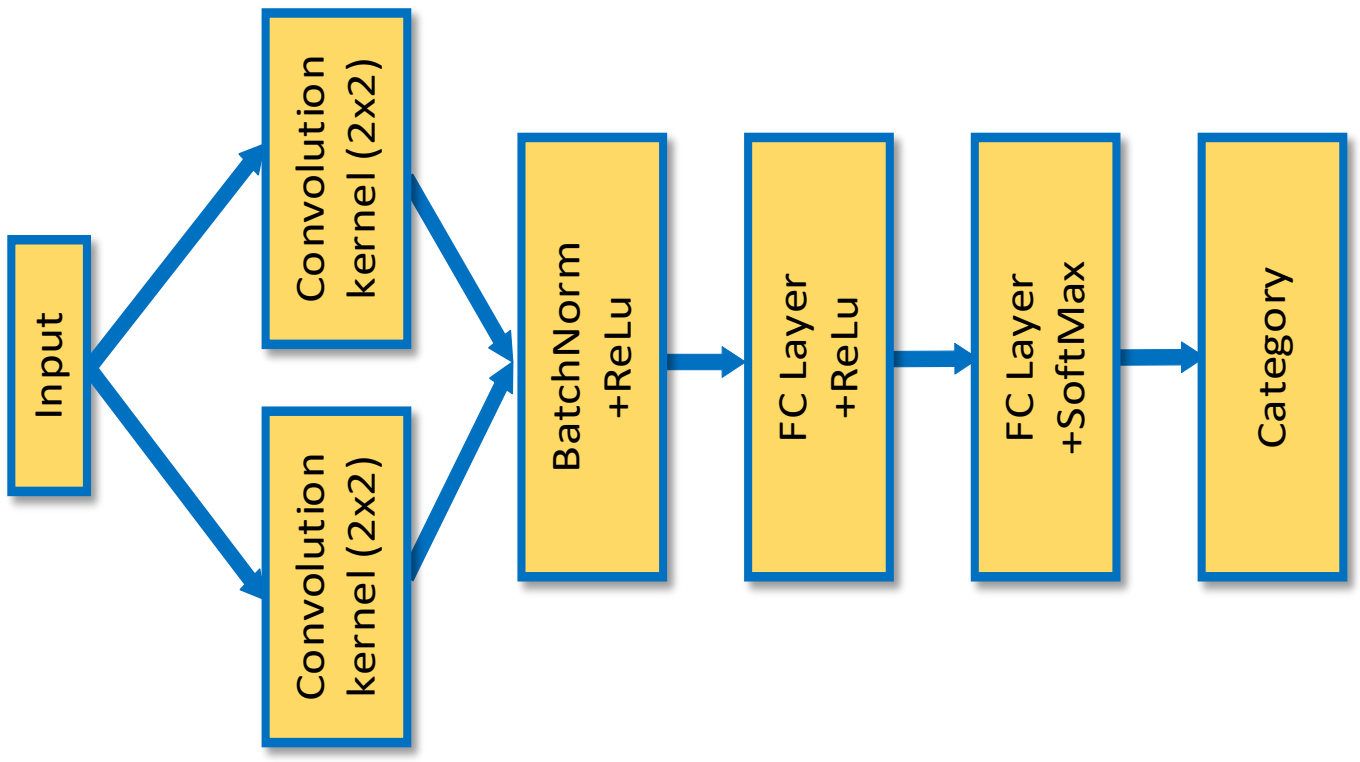

Fig. 17. The employed CNN model.

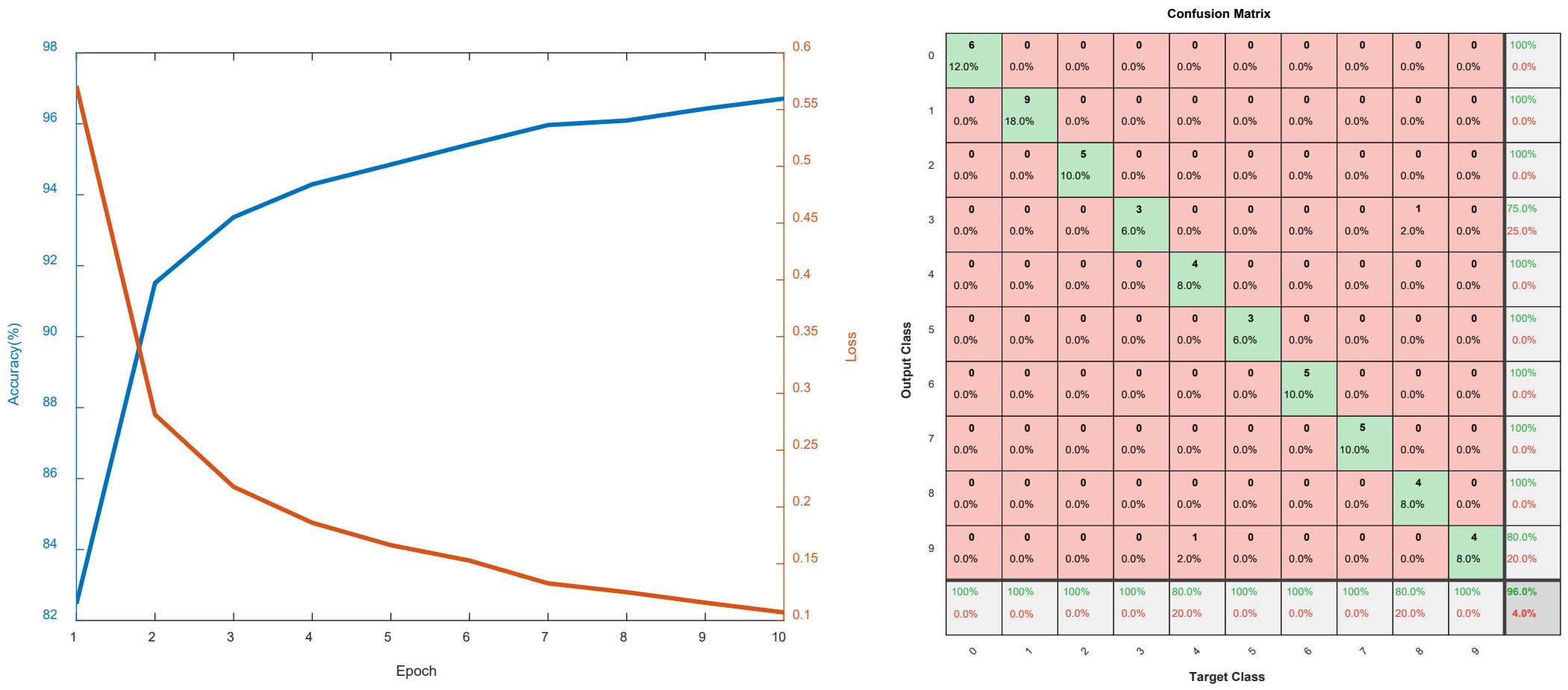

Fig. 18. (a) The accuracy and the cross-entropy loss during 10 epochs of training. (b) The confusion matrix of classifying 50 images in the MNIST test dataset.

Finally, the performance of the proposed photonic processor is analyzed. Firstly, the computational speed is evaluated. Suppose the circuit size of the photonic processor is $N\times N$, and then the computational speed $S$ can be given by

$$S \approx 4N^2 f_{clk} \tag{2}$$

where $f_{clk}$ is the clock frequency. In the experiment, the computational speed is $4\times2^2\times8$ GHz = 0.128 TOPS, where TOPS stands for "Tera Operations Per second". Fig. 19 shows the relationship between the circuit size $N$ and the computational speed under different frequency clocks. When the size of the photonic processor is 50×50 and the clock frequency is 8 GHz, the computational speed is 80 TOPS. If the clock frequency further increases to 16 GHz, the computational speed will increase to 160 TOPS.

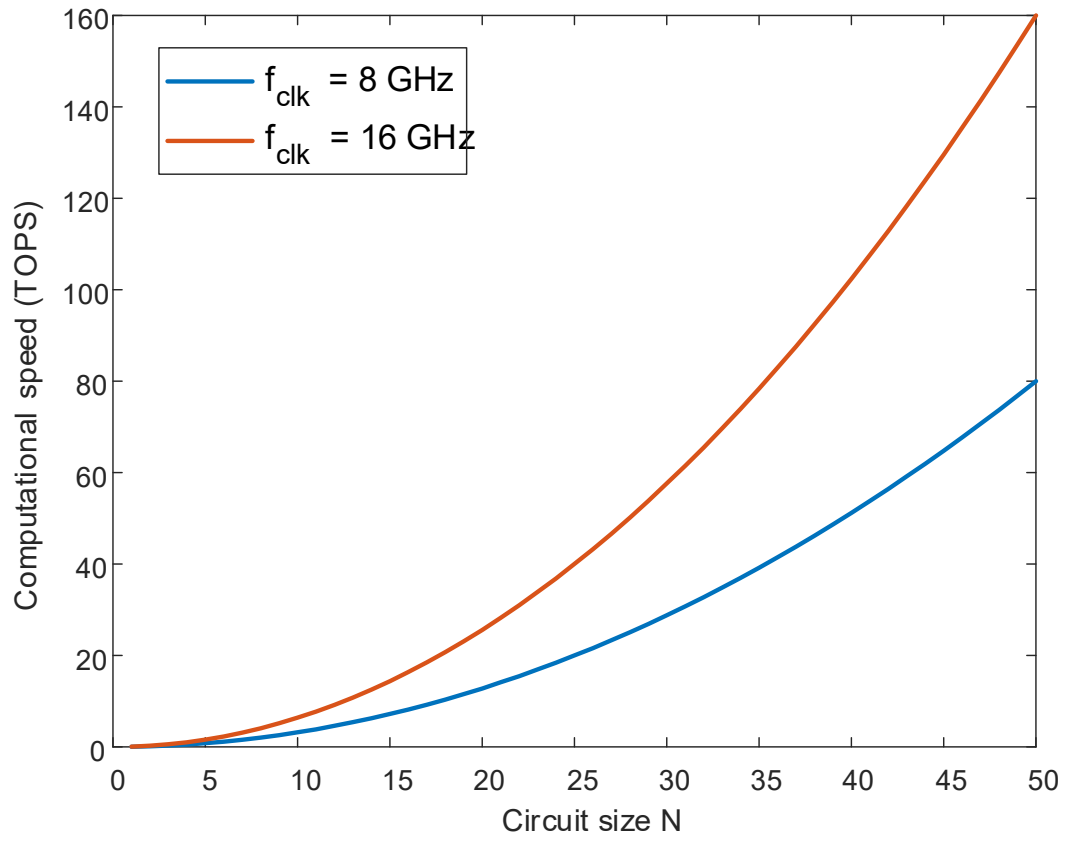

Fig. 19. Computational speed versus circuit size under different clock frequencies.

Then, the computational density of the photonic processor is evaluated. Suppose the length of each mesh is $L$. Then, the whole area of the photonic chip is about

$$A \approx N^2 L^2 \tag{3}$$

In the experiment, the used area of the photonic chip is about 0.0225 mm$^2$. If the circuit size increases to 50, the whole area will be 14.0625 mm$^2$.

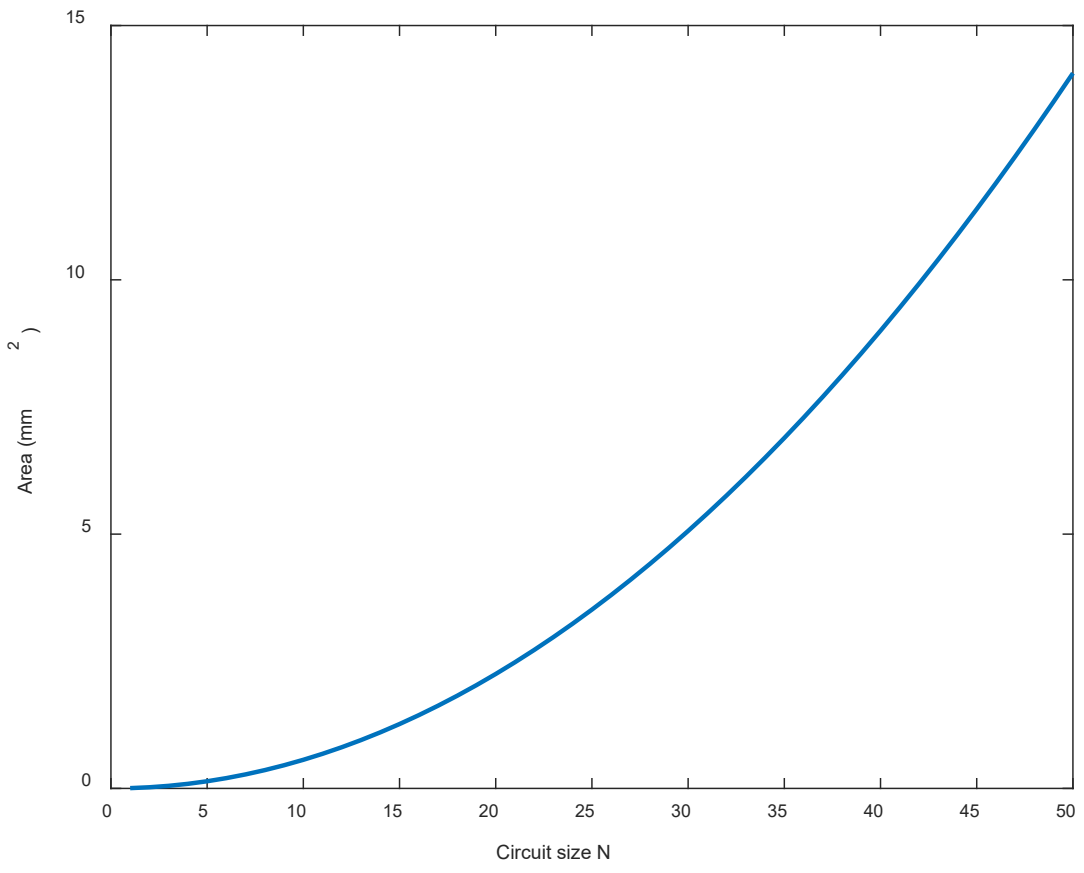

Fig. 20. Chip area versus circuit size.

Then, the power consumption of the chip is given by [24]

$$P \approx f_{clk} E_{signal} N + 2N^2 E_{ps} \tag{4}$$

where the first and second terms are the power consumptions related to the optical signal and heaters of the MDRs, respectively. In the first term, the power consumption related to sending and receiving optical signals per clock $E_{signal}$ is assumed to be 13.3 pJ/clock, which contains 8.8 pJ/clock for an analog-to-digital converter, 3.5 pJ/clock for a digital-to-analog converter, and 1pJ/clock for other systems of optical interconnection [24]. In the second term, $E_{ps}$ is the average power consumption of a heater. In the case of the typical TO heater, the average power consumption is 10 mW [24]. In the experiment, the power consumption is calculated to be 0.2928 W. When the circuit size increases to 50 and the clock frequency is 8 GHz, the power consumption is 55.32 W. If the clock frequency further increases to 16 GHz, the power consumption will be 60.64 W.

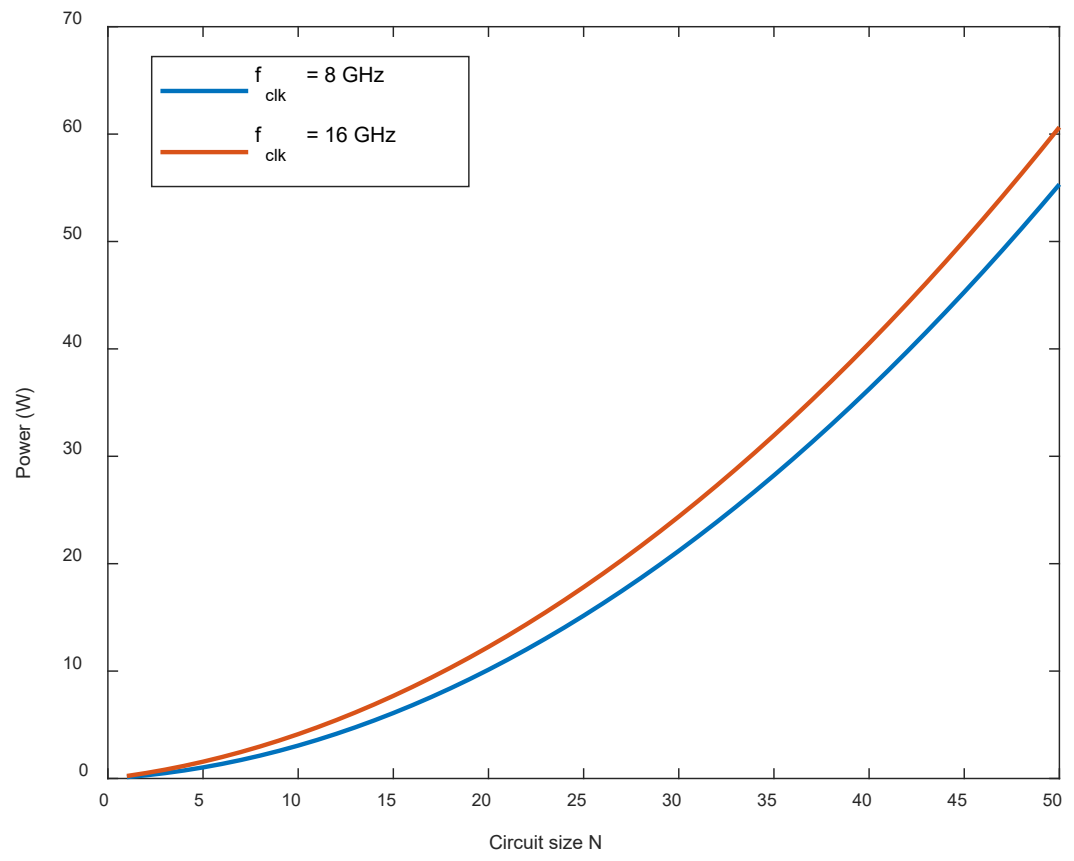

Fig. 21. Power consumption versus circuit size under different clock frequencies.

Given the power consumption and the chip area, the computational efficiency (CE) can be expressed as

$$\mathrm{CE} = P / S = E_{signal} / (4N) + E_{ps} / (2 f_{clk}) \tag{5}$$

The computational efficiency of the chip used in the experiment is calculated to be 2.2875 pJ/Operation. Fig. 22 shows the relationship between the computational efficiency and the circuit size under different clock frequencies. When the circuit size is 50 and the clock frequency is 8 GHz, the computational efficiency is 0.6915 pJ/Operation. When the clock frequency increases to 16 GHz, the computational efficiency can further improve to 0.379 pJ/Operation.

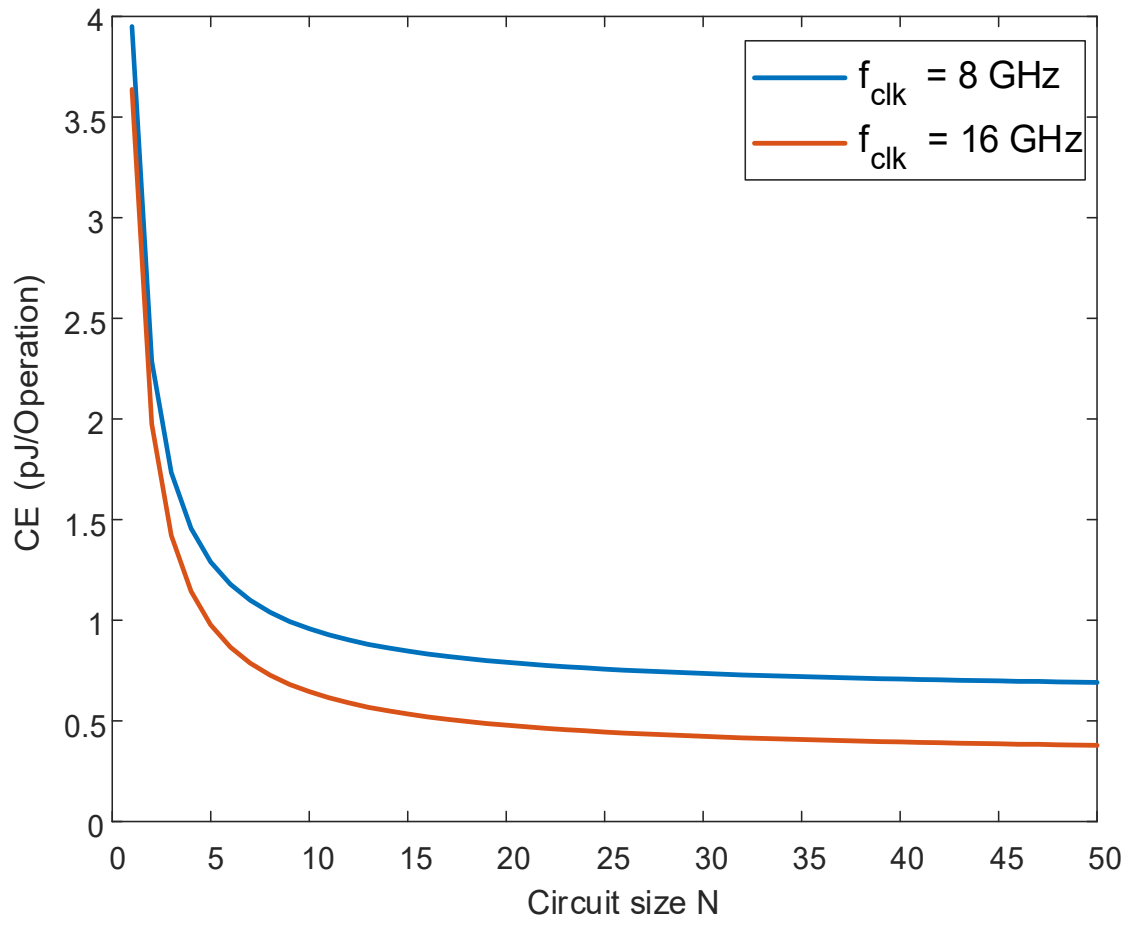

Fig. 22. Computational efficiency versus circuit size *N* under different clock frequencies.

Finally, the computational density (CD) of the chip can be given by

$$\mathrm{CD} = S / A = 4 f_{clk} / L^2 \tag{6}$$

The computational density of the utilized photonic processor is calculated to be 5.6889 TOPS/mm$^2$. Fig. 23 shows the relationship between the computational density and the cell length under different clock frequencies. As can be seen from Fig. 23, when the cell length decreases to 50 μm and the clock frequency is 8 GHz, the computational density is 12.8 TOPS/mm$^2$. If the clock frequency further increases to 16 GHz, the computational density will be 25.6 TOPS/mm$^2$.

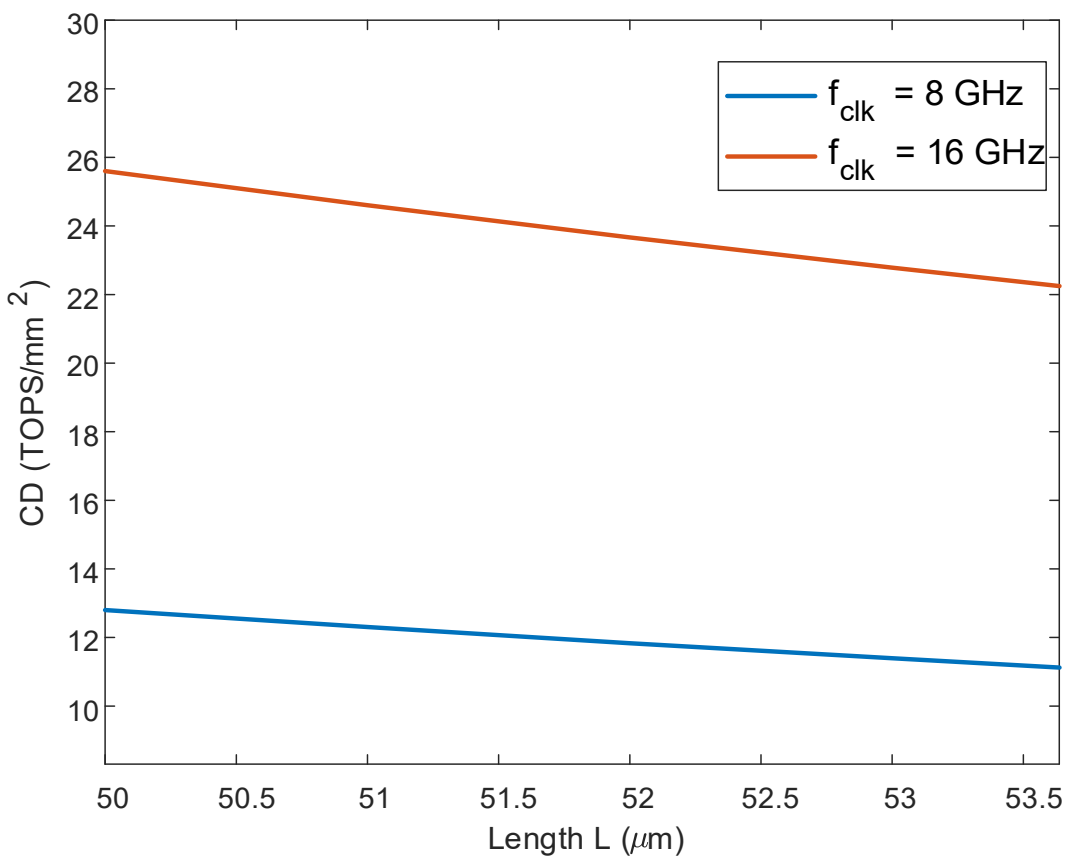

Fig. 23. Computational density versus cell length under different clock frequencies.

A comparison among the representative integrated photonic computing hardware is summarized in Table 1. Since two MDRs are placed at each cross of the crossbar, the computational density in this work is increased to 5.7 TOPS/mm$^2$. The overall computational density of the prototypical crossbar can be further improved by increasing the clock frequency and reducing the chip area by decreasing the mesh size of the crossbar.

**Table 1 Performance comparison of our proposed optical computing framework**

| **Technology** | **Matrix dimension** | **Platform** | **Task** | **Accuracy on the task** | **Efficiency (pJ/Operation)** | **Precision** | **Density (TOPS/mm$^2$)** |
|---|---|---|---|---|---|---|---|
| vector convolutional accelerator [26] | 9×10 | Discrete fiber optic components | MNIST | 88% | 0.79 | 7-bit | N/A |
| MZI [17] | 4×4 | Si | Vowel recognition | 76.70% | 0.015 | 5-bit | 1.12 |
| Micro-resonator banks [27] | 4×4 | Si | N/A | N/A | 0.09 | 4-bit | 3.20 |
| Micro-resonator banks [28] | 1×4 | Si | MNIST | 97.41% | 0.28 | 4-bit | 5.78 |
| PCM [23] | 9×4 | SiN | MNIST | 95.30% | 2.50 | 7-bit | 1.20 |
| PCM [29] | 1×4 | SiN | MNIST | 91.00% | N/A | 6-bit | 164.00 |
| IDNN [30] | 10×10 | Si | MNIST | 89.40% | N/A | N/A | N/A |
| Multimode interference [31] | 4×4 | SiN | MNIST | 92.17% | 2.42 | 5-bit | 25.48 |
| Micro-resonator crossbar [24] | 4×4 | Si | Iris dataset | 93% | N/A | N/A | N/A |
| Micro-resonator banks [32] | 2×2 | Si | MNIST | 96.6% | 7.71 | 9-bit | 0.136 |
| This work | 2×2×2 | Si | MNIST | 96% | 2.29 | 5.3-bit | 5.7 |

# Conclusion

In this work, we proposed a novel optical crossbar architecture to improve the computational density. Since each cross of the proposed crossbar has two MDRs which can function routing and weighting simultaneously, the computational density was doubled compared to the other crossbar architecture. We fabricated a silicon photonic 4×4 MDR crossbar, and experimentally demonstrated the use of the fabricated photonic processor for image classification in a CNN with two 2×2 kernels. Our approach to convolutional processing provides an effective method for removing the computing bottleneck in machine learning hardware.

# Method

Optical convolution computing with the proposed optical processor was implemented using commercially available optoelectronic components. The laser array is a Keysight N7714A laser source with four tunable polarization-maintaining output ports to generate four wavelengths. The four wavelengths are combined by an OC to form the OFC. The MZM used in the experiment has a bandwidth of 10 GHz. The AWG is Keysight M8195A with a sampling rate of 64 Sa/s. The waveform generated by the AWG is amplified by an electrical amplifier with a bandwidth of 15 GHz, and the amplified waveform is used to drive the MZM. The optical signal after the MZM is amplified by an Erbium-doped amplifier (EDFA) with a gain of 20 dB and then split into two branches. A tunable delay line is incorporated into one branch to realize 1 symbol time delay between the two branches. The PDs are used for optical-to-electrical conversion with a bandwidth of 10 GHz. The temporal waveforms are sampled by a real-time oscilloscope (Agilent DSO-X 93204A) with a sampling rate of 80 GSa/s.

**Acknowledgements**

The work is supported by the Natural Sciences and Engineering Research Council of Canada under the Silicon Electronic-Photonic Integrated Circuits (Si-EPIC) CREATE program. We acknowledge the CMC Microsystems, for providing the design tools and enabling the fabrication of the device.

**Author contributions**

L. H. and J. Y. conceived and designed the experiments, and L. H. performed the experiments and analyzed the data. L. H. and J. Y. wrote the paper.